\documentclass[10pt,letterpaper,twocolumn]{article} 

\usepackage{ol2}
\usepackage{amsmath}
\usepackage{amssymb}
\usepackage{multirow}
\usepackage[dvipdfm,colorlinks,linkcolor=blue,citecolor=blue,urlcolor=blue,anchorcolor=blue]{hyperref}

\begin{document}

\twocolumn[ 

\title{Optical vortices induced in nonlinear multi-level atomic vapors}


\author{Yiqi Zhang,$^{1,*}$ Zhenkun Wu,$^1$ Chenzhi Yuan,$^1$ Xin Yao,$^{1}$ Keqing Lu,$^2$ \\ Milivoj Beli\'c,$^{3}$ and Yanpeng Zhang$^{1,\dag}$}

\address{
$^1$Key Laboratory for Physical Electronics and Devices of the Ministry of Education \& Shaanxi Key Lab of Information Photonic Technique,
Xi'an Jiaotong University, Xi'an 710049, China \\
$^2$School of Information and Communication Engineering, Tianjin Polytechnic University, Tianjin 300160, China \\
$^3$Science Program, Texas A\&M University at Qatar, P.O. Box 23874 Doha, Qatar \\
$^*$Corresponding author: zhangyiqi@mail.xjtu.edu.cn \\
$^\dag$Corresponding author: ypzhang@mail.xjtu.edu.cn
}

\begin{abstract}
In a numerical investigation, we demonstrate the existence and curious evolution of vortices in a ladder-type
three-level nonlinear atomic vapor
with linear, cubic, and quintic susceptibilities considered
simultaneously with the dressing effect.
We find that the number of beads and topological charge of the incident beam, as well as its size,
greatly affect the formation and evolution of vortices.
To determine the number of induced vortices and the corresponding rotation direction,
we give common rules associated with the initial conditions coming from various incident beams.
\end{abstract}


 ] 

\noindent
An optical vortex is an interesting structure that possesses a phase defect at a point (called the vortex core)
and a rotational energy flow around it.
In recent decades, optical vortices and vortex solitons attracted a lot of attention from research groups all over the world,
for their potential applications in
optical data storage \cite{voogd_2004},
distribution \cite{tikhonenko_josab_95},
and processing \cite{crabtree_ao_2004}
as well as in the study of optical tweezers \cite{grier_nature_2003},
trapping and guiding of cold atoms \cite{kuga_prl_1997},
and entanglement states of photons \cite{mair_nature_2001}.
To date, research on optical vortices in many kinds of media, such as
bulk nonlinear media \cite{kivshar_2005},
discrete systems \cite{lederer_pr_2008},
atomic vapors \cite{skupin_prl_2007},
dissipative optical systems \cite{mihalache_prl_2006},
and Bose-Einstein condensates \cite{theocharis_prl_2003},
has been reported.
In Ref. \cite{michinel_prl_2006}, the authors discussed the properties of multidimensional beams via atomic coherence.
Still, there are many interesting topics to be researched.

In this Letter, we consider paraxial propagation of probe vortex beams
in a ladder-type three-level atomic system formed by $3S_{1/2}$, $3P_{1/2}$ and $5S_{1/2}$ levels of sodium.
The model is described by the nonlinear Schr{\"o}dinger equation (NLSE) of the form \cite{michinel_prl_2006,mihalache_prl_2006}:

\begin{equation}\label{initial_eq}
i\partial _z \psi  + \frac{1-i\beta}{{2k}}\nabla^2_\bot  \psi  + \frac{k}{2}\chi \psi  = 0,
\end{equation}
where $\nabla^2_\bot=\partial_{xx}+\partial_{yy}$ is the transverse Laplacian,
$k$ is the wave number, $\psi$ is the amplitude of the beam,
$\beta$ is the diffusion coefficient, and
$\chi$ is the total susceptibility of the atomic vapor system.
We consider a cubic-quintic NLSE as the underlying propagation model that adequately describes the physics of atomic vapor systems.
The total susceptibility can be obtained through the Liouville pathways using perturbation theory \cite{zhang_2009}:
$\chi = {\eta}/{K} -({\eta}/{K^2})(| G_1 |^2 /{d_1} + | G_2  |^2/{d_2} )+ ({\eta}/{K^3})( | {G_1 } |^4 /{d_1^2} + | {G_2 } |^4/{d_2^2 } + | {G_1 } |^2 | {G_2 } |^2 /{d_2^2 } )$,
with $\eta=iN\mu _{10}^2/(\hbar \varepsilon _0)$,
$d_1=\Gamma _{20}  + i\Delta _1 $,
$d_2=\Gamma _{20}  + i\left( {\Delta _1  + \Delta _2 } \right)$,
and
$K = \left( {\Gamma _{10}  + i\Delta _1 } \right) + {\left| {G_2 } \right|^2 }/d_2$,
where $G_1=\mu_{10} \psi/\hbar$ is the Rabi frequency of the probe field,
$G_2$ is the Rabi frequency of the coupled field,
$N$ is the atomic density, and
$\Delta_{1,2}$ is the detuning of the probe (coupled) field.
$\Gamma_{ij}$ denotes the population decay rate between the corresponding energy levels $|i\rangle$ and $|j\rangle$,
and $\mu_{10}$ is the electric dipole moment.
The second and the third terms in $\chi$ represent the cubic and the quintic contributions to the total susceptibility.
Equation (\ref{initial_eq}) is similar to the complex Ginzburg-Landau equation,
with the diffusion coefficient $\beta$ coming from the models of laser cavities.
We launch an incident beam in Eq. (\ref{initial_eq}) of the form (in polar coordinates):
\begin{equation}\label{azimuthon}
    \begin{split}
      \psi \left( {z=0,r,\theta } \right) =& A{\rm{sech}}\left[ \left(r - R_0\right)/r_0 \right] \\
        &  \times [\cos(n\theta)+iB\sin(n\theta)] \exp(il\theta)
    \end{split}
\end{equation}
with $B=0,~n\neq 0$ for a necklace, $B\neq0,~n\neq0$ for an azimuthon \cite{zhang_oe_2010}, and $B=1$ or $n=0$ for a vortex.
Here $R_0$ is the mean radius, $r_0$ is the width,
$A$ is the amplitude, $B$ is the modulation coefficient ($1-B$ is the modulation depth),
$l$ is the input topological charge, and $2n$ is the number of necklace beads.
The beam in Eq. (\ref{azimuthon}) can be viewed as a superposition of two vortices
with net topological charges (NTCs) \cite{molina_josab_2000} $l+n$ and $l-n$, respectively.
The vortex with a larger magnitude of NTC will dissipate faster during propagation,
due to the diffusion term in Eq. (\ref{initial_eq}) \cite{he_oe_2007}.
Thus, the vortex with a smaller NTC will remain stable for longer during propagation,
and the overall NTC will correspond to that charge.

\begin{figure}[htbp]
\centering
\includegraphics[width=\columnwidth]{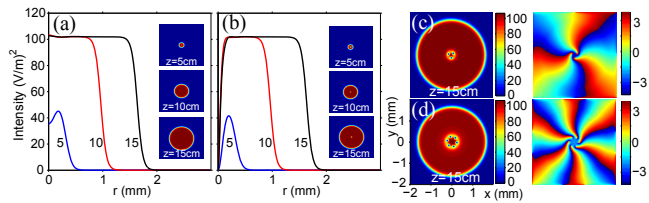}
\caption{(color online) (a) and (b) Evolution of vortex incidences,
for several propagation distances and for
$l=0$ and $l=1$, respectively.
(c) and (d) Output intensities (left panels) and phases (right panels) of vortices with $l=3$ and
$l=6$, respectively.
The colorbars and physical scales of the transverse plane shown here are the same in all other figures.
}
\label{fig1}
\end{figure}

First of all, we set the parameters to be
$N=10^{13}~{\rm{cm}}^{-3}$,
$\mu_{10}=3\times 10^{-29}~{\rm{Cm}}$,
$\Delta_1=1~{{\rm{MHz}}}$,
$\Delta_2=-1~{{\rm{MHz}}}$,
$G_2=40~{{\rm{MHz}}}$,
$\Gamma_{10}=2 \pi \times4.86~{{\rm{MHz}}}$,
$\Gamma_{20}=2 \pi \times0.485~{{\rm{MHz}}}$,
$\lambda=600~{\rm{nm}}$,
$r_0=100~\mu{\rm{m}}$,
$R_0=200~\mu{\rm{m}}$,
and $\beta=0.5$ in our numerics for convenience without special statement \cite{steck_sodium}.
The evolution of a vortex with $l=0$ is shown in the insets in Fig. \ref{fig1}(a).
The input notch of the vortex soon disappears and
the vortex changes into a super-Gaussian-like pulse during propagation,
with the width and amplitude growing dramatically.
For comparison, if we set $l=1$ and redo the evolution,
the width and the amplitude still grow,
but the notch at the origin remains, as shown in the insets in Fig. \ref{fig1}(b).
Even though the amplitude increases during propagation,
there exists a saturable maximum at $\sim 10~{\rm{V/m}}$,
as seen from the radial intensity profiles in Figs. \ref{fig1}(a) and (b);
these results are quite similar to the results from Refs. \cite{kivshar_2005} and \cite{michinel_prl_2006}.
The saturation phenomenon can be explained by the competition
between the cubic and the quintic nonlinearity \cite{zhang_oc_2012}.
The nonlinearity is defocusing, and as such can support stable vortices \cite{kivshar_2005}.
The beam spreads during propagation, because the nonlinearity is too weak to balance diffraction and form a soliton.
If $l$ is set to 3 and 6,
there will be 3 and 6 notches around the origin,
as shown in Figs. \ref{fig1}(c) and (d).
The right panels there present the corresponding phases that demonstrate every notch is a vortex.

\begin{figure}[htbp]
\centering
\includegraphics[width=\columnwidth]{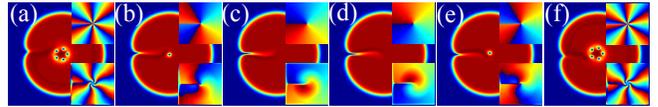}
\caption{(color online)
(a)-(f) Evolution outputs at $z=15~\rm{cm}$, from a crescent input ($n=0.5$)
corresponding to $l=-5.5,~-1.5,~-0.5,~0.5,~1.5$, and $5.5$, respectively.
The top and bottom insets present the input and output phases, respectively.
}
\label{fig2}
\end{figure}

If we set $n=0.5$, we can investigate the evolution of a unique ``necklace'' with crescent shape.
In Fig. \ref{fig2} we exhibit a series of evolution snapshots corresponding to different $l$.
By comparing Figs. \ref{fig2}(b) with (e), and (c) with (d),
one can conclude that the number of notches that appear in the beam is determined by $\min\{|l\pm n|\}$,
which is also the absolute value of the NTC of the survived vortex component.
Even though Figs. \ref{fig2}(a) and (f) have the same number of vortices,
they rotate in opposite directions,
because the corresponding NTCs are $-5$ and $5$, respectively.
The common rule on how to calculate the number of vortices
and how to determine their rotation directions
is presented in Table \ref{table1},
where $\circlearrowleft$ and $\circlearrowright$ represent rotation senses of the vortices.

\begin{table}[htbp]
  \centering
  \caption{Properties of the induced vortices from necklace and azimuthon incidences with different $l$ and $n$.}
  \newsavebox{\tablebox}
  \begin{lrbox}{\tablebox}
  \begin{tabular}{*{9}{c}} \\ \hline
\multicolumn{8}{l}{(1) Necklace inputs}  \\ \hline
& $l < -n$ & $l=-n$ & $-n < l < 0$ & $l=0$ & $0<l<n$ & $l=n$ & $l>n$  \\ \hline
& $\circlearrowright$ & - & $\circlearrowleft$ & - & $\circlearrowright$ & - & $\circlearrowleft$ \\
No. & $-n-l$ & 0 & $n+l$ & 0 & $n-l$ & 0 & $-n+l$ \\ \hline \hline
\multicolumn{8}{l}{(2) Azimuthon inputs when $r_0>r_{cr1}$}  \\ \hline
& $l < -n$ & $l=-n$ & $-n < l < 0$ & $l=0$ & $0<l<n$ & $l=n$ & $l>n$  \\ \hline
outer & - & - & - & - & $\circlearrowleft$ & $\circlearrowleft$ & $\circlearrowleft$ \\
inner & $\circlearrowright$ & - & $\circlearrowleft$&  $\circlearrowleft$& $\circlearrowright$ & - & $\circlearrowleft$ \\
No. & $-n-l$ & 0 & $n+l$ & $n$ & $3n-l$ & $2n$ & $n+l$ \\ \hline \hline
\multicolumn{8}{l}{(3) Azimuthon inputs when $r_0<r_{cr2}$ with $l>n$}  \\ \hline
& \multicolumn{2}{c}{$n=1$} & \multicolumn{3}{c}{$n=2~(l\leq5|l>5)$} &\multicolumn{2}{c}{$n\geq3$} \\ \hline
outer & \multicolumn{2}{c}{$\circlearrowleft$} & \multicolumn{3}{c}{$\circlearrowleft|\circlearrowleft$} & \multicolumn{2}{c}{-} \\
inner & \multicolumn{2}{c}{$\circlearrowleft$} & \multicolumn{3}{c}{$\circlearrowleft|\circlearrowleft$} & \multicolumn{2}{c}{$\circlearrowleft$} \\
No. & \multicolumn{2}{c}{$l+n$} & \multicolumn{3}{c}{$l+n|l-n$} & \multicolumn{2}{c}{$l-n$} \\ \hline
  \end{tabular}
  \label{table1}
  \end{lrbox}
  \scalebox{0.7}{\usebox{\tablebox}}
\end{table}

\begin{figure}[htbp]
\centering{\includegraphics[width=\columnwidth]{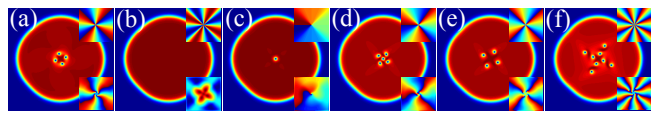}}
\caption{(color online)
Evolutions of azimuthon incidences with $n=2$, and $B=0.5$ for
$l=-6$ (a), $-2$ (b), $-1$ (c), 1 (d), 2 (e), and 6 (f), respectively.
The figure setup is as in Fig. \ref{fig2}.
}
\label{fig3}
\end{figure}

Numerical experiments indicate that there are two critical values
$r_{cr1}\approx 42.5 ~\mu m$ and $r_{cr2}\approx 36.1 ~\mu m$ for the beam width $r_0$.
The evolution of quadrupole azimuthons with $n=2$ and different $l$
corresponding to $r_0>r_{cr1}$ are shown in Fig. \ref{fig3}.
Following the rule from above,
the number of notches at the origin is determined by NTC;
so, from (a) to (f) in Fig. \ref{fig3},
the number of notches is 4, 0, 1, 1, 0, and 4, respectively.
However, in (d)-(f),
there are additional 4 notches around the origin, which equal the number of beads.
The explanation is that there is energy flow around the phase singularities,
and at the same time the beads fuse due to the diffusion term $\beta$;
therefore when the speed of energy flow is greater than the fusion speed of beads,
at the twist of the beads new vortices form.
It is worth mentioning that the vortices around the origin appear as vortex pairs (with charges of $+1$ and $-1$),
so the NTC of the beam is still conserved.
The common rule for this case is also shown in Table \ref{table1}, under the ``Azimuthon inputs when $r_0>r_{cr1}$'' case.

\begin{figure}[htbp]
\centering{\includegraphics[width=\columnwidth]{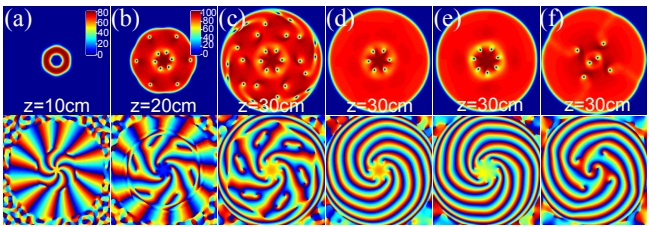}}
\caption{(color online)
(a)-(c) Evolution of an azimuthon with $l=9,~n=3,~B=0.5$, $r_0=40~\mu {\rm{m}}$ at different propagation distances.
(d)-(f) show azimuthons with $r_0=30~\mu {\rm{m}}$ at 30 cm
with $(l=9$, $n=3)$, $(l=9$, $n=2)$, and $(l=4$, $n=2)$, respectively.
The top row are the intensities, the bottom row are the phases.}
\label{fig4}
\end{figure}

A hexapole azimuthon with $l>n$ is presented in Figs. \ref{fig4}(a)-(c); its
$r_0=40~\mu{\rm{m}}$ is chosen to fall in-between $r_{cr2}$ and $r_{cr1}$. According to the rule from Table \ref{table1},
the total number of induced vortices should equal $n+l=12$.
However, from Fig. \ref{fig4} we see that there are more and more vortices appearing during propagation.
The reason is that the speed of energy flow is the largest, as compared to those of fusion and spreading.
From the phases at different propagation distances,
we see that the energy flow brought by the vortices around the origin of the fused beam
is always faster than that at the edge of the beam;
therefore, the asynchrony preferably forms new phase singularities at the edge of the beam,
and new vortices are induced correspondingly.
But we cannot give a certain rule for this case, because the number of induced vortices is greatly affected
by the beam width. What one sees in Fig. \ref{fig4}(c) is 6 vortices at the core plus 12 pairs of $\pm1$
vortices induced at the rim of the beam.

If we set $r_0=30~\mu \textrm{m}$, which fulfills the condition $r_0<r_{cr2}$,
and redo the propagation of the beam used in Figs. \ref{fig4}(a)-(c),
we find that the number of induced vortices is 6, which can be calculated from $l-n=6$,
as shown by the output intensities and phases in Fig. \ref{fig4}(d).
The common rule for this case is exhibited in Table \ref{table1} under
``Azimuthon inputs when $r_0<r_{cr2}$ with $l>n$''
(other cases are the same as those under ``Azimuthon inputs when $r_0>r_{cr1}$'').
The lower the number of beads,
the more energy in each of the beads, which will strengthen the energy flow.
That is why the number of induced vortices is $l+n$ if $n=1$.
However, the exact rule for $n=2$ is not certain, because the energy flow is weakened,
and whether it can produce more vortices or not, it is up to the value of $l$ set for the initial beam.
If $l$ is bigger, the fusion will be accelerated,
so the production of new vortices will be limited.
Here, $l=5$ is a boundary for this case: the rule is $l-n$ for $l>5$ and $l+n$ for $l\leq5$.
Figures \ref{fig4}(e) and (f) display two numerical experiments corresponding to $l=9$ and $l=4$,
in which the number of vortices are $l-n=7$ and $l+n=6$, respectively.

In conclusion, we have demonstrated that optical vortices can form
from vortex, necklace, and azimuthon incidences with different topological charges,
in multi-level atomic vapors when linear, cubic and quintic susceptibilities are considered simultaneously.
The appearance of vortices results from a combined action of the number of topological charges, the
beam width of incidences, the diffusion effect, cubic-quintic nonlinearities, and the loss/gain in the medium, simultaneously.
We have formulated common rules of finding the number as well as the rotation direction of the induced vortices.
The effects can be observed in sodium as well as other
atomic vapors with incident beams produced by using
multi-wave interference \cite{masajada_oc_2001} or phase mask \cite{chen_prl_1997} methods
in a similar experimental setup as that in Ref. \cite{sang_ieee_2012}.
Our findings open a new venue and introduce a new method for studying vortices,
and may broaden the field of their applications.

This work was supported by the 973 Program (2012CB921804),
NNSFC (10974151, 61078002, 61078020, 11104214, 61108017, 11104216),
NCET (08-0431), RFDP (20110201110006, 20110201120005, 20100201120031),
and FRFCU (2011jdhz07, xjj2011083, xjj2011084, xjj20100151, xjj20100100, xjj2012080).
Support from Qatar National Research Fund NPRP 09-462-1-074 project is also acknowledged.
Yiqi Zhang appreciates discussions with Dr. Ruimin Wang.

\end{document}